\documentclass[a4paper]{jpconf}

\usepackage[pdftex]{graphicx}
\usepackage[noadjust]{cite}
\usepackage{amssymb}
\usepackage{amsmath}
\usepackage{amsthm}
\usepackage{color}
\usepackage{float}
\usepackage[leftcaption]{sidecap}
\usepackage{comment}

\newcommand{\be}{\begin{eqnarray}}
\newcommand{\ee}{\end{eqnarray}}
\newcommand{\bez}{\begin{eqnarray*}}
\newcommand{\eez}{\end{eqnarray*}}
\newcommand{\bbR}{\mathbb{R}}
\newcommand{\cU}{\mathcal{U}}

\theoremstyle{plain}
\newtheorem{thm}{Theorem}[section]

\newtheorem{proposition}[thm]{Proposition}

\theoremstyle{definition}

\theoremstyle{remark}

\newtheorem{remark}[thm]{Remark}
\newtheorem{example}[thm]{Example}

\begin{document}

\title{KdV soliton interactions: a tropical view} 

\author{A Dimakis$^1$ and F M\"uller-Hoissen$^2$}

\address{$^1$ Dept. of Financial and Management Engineering, 
 Univ. of the Aegean, 82100 Chios, Greece}

\address{$^2$ Max-Planck-Institute for Dynamics and Self-Organization,
 37077 G\"ottingen, Germany}

\ead{dimakis@aegean.gr, folkert.mueller-hoissen@ds.mpg.de}

\begin{abstract}
Via a ``tropical limit'' (Maslov dequantization), Korteweg-deVries (KdV) solitons correspond to piecewise 
linear graphs in two-dimensional space-time. We explore this limit. 
\end{abstract}

\section{Introduction}
\setcounter{equation}{0}
The ``tropical limit'' (Maslov dequantization) of soliton solutions of the (scalar, real) 
KdV equation 
\bez
       4 \, u_t = u_{xxx} + 6 \, u \, u_x \, , 
\eez
where a subscript indicates a partial derivative with respect to an independent variable,
describes them as piecewise linear graphs in two-dimensional space-time. The corresponding 
exploration in this work is based on techniques developed in \cite{DMH11KPT,DMH12KPBT}, 
but our presentation will be fairly self-contained. 

The most striking property of KdV solitons 
is surely the well-known fact that, after an interaction, they regain their 
``identity'' (amplitude, width and speed) and only experience a shift in space. 
But what really happens during the interaction of solitons? After the ground-breaking work of Kruskal 
and Zabusky \cite{Zabu+Krus65}, this question has been addressed again and again  
\cite{Lax68,Bowt+Stua80,Bowt+Stua83,Yone84a,Yone84b,Caen+Malf85,Malf+VdVelde85,Molo+Hodn86,Hodn+Molo89,Molo+Hodn89b,Fuchss87,LeVe87,Brya+Stua92,Camp+Park97,Nguy03,Nguy04,BKY06,Bona+Weiss09}. 
Some authors argued that solitons always pass through one another (see, e.g., \cite{Hodn+Molo89,LeVe87}). 
Others suggested that solitons exchange their identities during the interaction 
(see, e.g., \cite{Bowt+Stua80,Bowt+Stua83}). 
Based on a certain decomposition of the $2$-soliton solution, some proposed an intermediate wave that
transfers energy between the two solitons \cite{Brya+Stua92,BKY06,Bona+Weiss09}.
There have been several attempts to ``individualize'' the a priori asymptotically (i.e., for $t \ll 0$) 
defined solitons also during an interaction. Solitons do not really behave like classical particles, 
however. As localized disturbances of a continuous medium, they possess a wavelike nature. 
Instead of speaking of the same incoming ($t \ll 0$) and outgoing ($t \gg 0$) soliton, 
it is more adequate to speak of ``instances'' of a certain soliton state. Once solitons start to 
interact, they loose their individuality.  
This is also what the tropical limit shows: two KdV solitons 
(which can be regarded as parallel KP line solitons, see \cite{DMH11KPT}, Example~4.5) interact 
by exchanging a ``virtual soliton'', see Section~\ref{sec:2solitons}. 
The present work explores more generally KdV soliton interactions in the tropical limit.
The analogy with a quantum scattering theory (also see \cite{Nguy03,Nguy04,BKY06}) is striking.  
But here a kind of second quantization is not necessary since the KdV equation (and any other evolution equation 
possessing solitons) is already a many-``particle'' model. Solitons are the asymptotically free particles, 
and their interaction can be understood as an exchange of virtual particles. We should stress, 
however, that we are \emph{not} attempting to interprete KdV as a quantum theory. 

Section~\ref{sec:KdV_solitons} expresses the KdV soliton solutions in a form convenient 
for our purposes. Section~\ref{sec:KdV_tropical} then deals with the tropical limit.

\section{KdV solitons}
\label{sec:KdV_solitons}
\setcounter{equation}{0}
Let us recall (see, e.g., \cite{Hiro04}) that the $M$-soliton solution of the KdV equation is given 
by $u = 2 \, (\log \tau_H)_{xx}$ with the Hirota $\tau$-function
\bez
    \tau_H = \sum_{\mu_1,\ldots,\mu_M = 0,1} \exp\Big( \sum_{j=1}^M \mu_j \eta_j
               + \sum_{1 \leq j < k \leq M} b_{jk} \, \mu_j \mu_k \Big) \, , 
\eez
where $\eta_j = 2 p_j \, (x + p_j^2 \, t + \tilde{c}_j)$ and 
$b_{jk} = \log [(p_k-p_j)^2/(p_k+p_j)^2]$.
This involves arbitrary real constants $\tilde{c}_j$ and $0 < p_1 < p_2 < \cdots < p_M$.

\begin{proposition}
An equivalent expression for the $M$-soliton KdV solution is 
\bez
 && \tau = \sum_{A \in \{-1,1\}^M} e^{\Theta_A} \qquad \qquad \mbox{where} \\
   \Theta_A = \sum_{j=1}^M \alpha_j \, \theta_j + \delta_A \, , 
 &&  \theta_j = p_j \, ( x + p_j^2 \, t + c_j) \, , \quad
   \delta_A = \log\Delta_A \, , \quad
   \Delta_A = |\Delta(\alpha_1 p_1,\ldots,\alpha_M p_M)| \, ,
\eez 
with $A=(\alpha_1,\ldots,\alpha_M)$, $\alpha_j \in \{\pm 1\}$, and $\Delta$ is a Vandermonde determinant.
\end{proposition}
\noindent
\textit{Proof.} 
With $\tau$ we associate the equivalent (no change in $u$) tau function
\bez
    \tau' = \frac{1}{\Delta_{(-1,\ldots,-1)}} \, e^{\theta_1 + \cdots + \theta_M} \, \tau 
           = \sum_{\alpha_1,\ldots,\alpha_M=\pm 1} 
             \frac{\Delta_{(\alpha_1,\ldots,\alpha_M)}}{\Delta_{(-1,\ldots,-1)}} \, 
             e^{(\alpha_1+1) \theta_1 + \cdots + (\alpha_M+1) \theta_M}
            \; .
\eez
With the help of 
\bez
    \Delta_{(\alpha_1,\ldots,\alpha_M)} = \prod_{1\leq j < k \leq M} (p_k - \alpha_j \alpha_k \, p_j)
             = \prod_{1\leq j < k \leq M} (p_k - p_j)^{(1 + \alpha_j \alpha_k)/2}
               (p_k + p_j)^{(1 - \alpha_j \alpha_k)/2} \, ,
\eez
we find that $\tau' = \tau_H$ if we set \\
\hspace*{.3cm}
\begin{minipage}{10cm}
\bez
    \mu_j := (\alpha_j +1)/2 \in \{0,1\} \, , \qquad
    \tilde{c}_j := c_j + \frac{1}{2 p_j} \log \Big| \prod_{k \neq j} \frac{p_k+p_j}{p_k-p_j} \Big| \; . 
\eez
\end{minipage}
\hfill \begin{minipage}{.6cm} \vspace*{.8cm} $\square$\end{minipage}

\vspace{.4cm}
For an $M$-soliton solution, there are $2^M$ different phases $\Theta_A$. 
For $M>2$ we extend the variables $x$ and $t$ to $N \geq M$ (KdV hierarchy) variables $t^{(k)}$, $k=1,\ldots,N$, 
where $x = t^{(1)}$, $t = t^{(2)}$. Via translations of these variables, we 
achieve that $c_j =0$, $j=1,\ldots,N$, so that 
\be
     \theta_j = \sum_{k=1}^N p_j^{2k-1} \, t^{(k)} \; .  \label{theta_redef}
\ee

\begin{remark}
\label{rem:reflection}
After having arranged (\ref{theta_redef}) by a shift of the KdV hierarchy variables, 
an $M$-soliton solution is invariant under reflection of 
all independent variables, i.e., $t^{(k)} \mapsto -t^{(k)}$, $k=1,\ldots,N$. 
\end{remark}

\begin{example}
For the $2$-soliton solution, setting $N=2$, we obtain
\bez
    u(x,t) = 8 (p_2^2 -p_1^2) \frac{p_1^2 \, \sinh^2 \theta_2 + p_2^2 \, \cosh^2 \theta_1}{[(p_2-p_1) \, \cosh(\theta_1+\theta_2)
        + (p_2+p_1) \, \cosh(\theta_1-\theta_2)]^2} \; .
\eez
Since we set $c_1=c_2=0$, we have $u(x,t)=u(-x,-t)$, so that the solution is adapted to an obvious 
symmetry of the KdV equation. In particular, $u(x,0)$ is symmetric about 
$x=0$ (also see Theorem 1 in \cite{Bona+Weiss09}, where the expressions for the symmetry event are 
more complicated due to a different form of the KdV equation and a different parametrization of the 
$2$-soliton solution). We find 
$u_{xx}(0,0) = - 4 (p_2-p_1)(p_2+p_1)(p_2^2 - 3 p_1^2)$. 
Hence $u(x,0)$ has a minimum at $x=0$ if $p_2/p_1 < \sqrt{3}$ and a maximum 
if $p_2/p_1 > \sqrt{3}$ \cite{Lax68,Malf+VdVelde85,Bona+Weiss09}, and we find  
$u(0,0)= 2 (p_2^2 - p_1^2)$.  
\end{example}

\section{Tropical limit of KdV solitons}
\label{sec:KdV_tropical}
\setcounter{equation}{0}
Let $\cU_B$ be the region in $\bbR^2$ where the phase $\Theta_B$ dominates all others, i.e.,
\bez
    \cU_B = \left\{ (x,t) \in \bbR^2 \, | \, \max\{ \Theta_A(x,t) \, | \, A \in \{-1,1\}^M \} 
            = \Theta_B(x,t) \, \right\} \, ,
\eez
considered at fixed values of the higher KdV hierarchy variables. 
$\cU_B$ is connected, it may be empty for some $B$. 
In a non-empty set $\cU_B$ and sufficiently far away from the boundary, the approximation
\bez
     \log \tau \simeq \max\{ \Theta_A \, | \, A \in \{-1,1\}^M \}
\eez 
is valid, 
so that $u$ vanishes (since $\Theta_B$ is linear in $x$). As a consequence, $u$ is localized along the 
boundary lines of non-empty dominating phase regions. The \emph{tropical limit of the KdV soliton solution} 
is the piecewise linear planar graph consisting of such boundary lines, and the amplitude $u$ on these lines. 
The structure of this graph is determined by the intersections of the dominating phase regions. 
In the following, we write $\cU_{AB} := \cU_A \cap \cU_B$. 

\begin{remark}
A precise formulation of what we call the tropical limit of KdV solitons is obtained by using the 
Maslov dequantization formula (see, e.g., \cite{Litv10})
\be
     \lim_{\epsilon \to 0} \; \epsilon \sum_{A \in \{-1,1\}^M} e^{\Theta_A/\epsilon} 
    = \max\{ \Theta_A \, | \, A \in \{-1,1\}^M \} \, ,   \label{Maslov_deq}
\ee
which replaces the operation of addition (of exponentials) by the maximum function 
(applied to the phases). This is a familar step in ``tropicalization''. 
It is usually accompanied by also replacing multiplication by addition (the ``tropical product''). 
\end{remark}

\begin{remark}
If the logarithmic terms $\delta_A$ were negligible, 
then we would have $\log \tau \simeq \max_{A \in \{-1,1\}^M}\{ |\sum_{j=1}^M \alpha_j \theta_j|\} 
= \sum_{j=1}^M |\theta_j|$,
and the tropical limit of the KdV soliton would be given by the \emph{superimposition} of 
the space-time lines corresponding to the constituent single solitons. 
(\ref{Maslov_deq}) shows that this simplified limit 
corresponds to introducing ``slow variables'' via $x \mapsto x/\epsilon$ and $t \mapsto t/\epsilon$ 
in a soliton solution. This simply maps the latter to the corresponding solution of   
the $\epsilon$-dependent KdV equation $u_t - (3/2) \, u \, u_x = \epsilon^2 \, u_{xxx}$,
which formally approaches its ``dispersionless'' (or ``quasiclassical'') limit, the 
inviscid Burgers (or Hopf, or Riemann) equation $u_t - (3/2) \, u \, u_x =0$, as $\epsilon \to 0$. 
The $\epsilon$-dependent soliton solution does \emph{not} tend to a solution of the 
inviscid Burgers equation, however.
In the limit, KdV solitons ``disappear'' in the sense that their support becomes a set 
of measure zero in space-time, while $u$ retains a finite value. The associated initial data become 
infinitely steep, so there is no corresponding local solution of the inviscid Burgers equation 
via the method of characteristics. It is also not adequate to think of the (simplified) tropical 
limit as a kind of non-smooth solution of the inviscid Burgers equation (e.g., similar to those obtained 
via ``front tracking'' \cite{Hold+Rise11}). This is so because irrespective how small $\epsilon$ is, 
the corresponding $\epsilon$-dependent solution necessarily retains dispersion since it 
remains solitonic. 
We have to conclude that the tropical limit cannot be regarded as a dispersionless limit. 
\end{remark}

Introducing
\bez
    p_A := \sum_{j=1}^M \alpha_j \, p_j \, , \quad
    q_A := \sum_{j=1}^M \alpha_j \, p_j^3  \, , \quad
    c_A := \delta_A + \sum_{k=3}^N \sum_{j=1}^M \alpha_j \, p_j^{2k-1} \, t^{(k)}  \, , 
\eez
where $A=(\alpha_1,\ldots,\alpha_M)$, we have
\be
    \Theta_A - \Theta_B = (p_A -p_B) \, x + (q_A - q_B) \, t + c_A - c_B \; .  \label{phase_diff0}
\ee
There is a line at which two phases $\Theta_A$ and $\Theta_B$ ``meet'', 
i.e., where $\Theta_A = \Theta_B$. It is given by
\be
    x = x_{AB}(t) := - \frac{q_A-q_B}{p_A-p_B} \, t - \frac{c_A-c_B}{p_A-p_B}  \, ,
   \label{boundary_line}
\ee
assuming that $p_A \neq p_B$, and (\ref{phase_diff0}) can be written as
\be
    \Theta_A - \Theta_B = (p_A -p_B) (x - x_{AB}(t)) \; .   \label{phase_diff}
\ee
If $(x,t) \notin \cU_B$, there is a phase $\Theta_A$ such that $\Theta_A(x,t) > \Theta_B(x,t)$. 
In this case we say that the phase $\Theta_B$ is \emph{non-visible} at the event $(x,t)$. 
An immediate consequence of the last identity is
\bez
     p_A > p_B \quad \Longrightarrow \quad 
     \begin{array}{l}
     \Theta_B(x,t) \\  
     \Theta_A(x,t) 
     \end{array} \quad
     \mbox{is non-visible for} 
     \quad
     \begin{array}{l} x > x_{AB}(t) \\
                      x < x_{AB}(t) \; .
     \end{array} 
\eez
Near an event on the boundary line (\ref{boundary_line}) that is not a higher order coincidence 
of phases, we have $\tau \simeq e^{\Theta_A} + e^{\Theta_B}$ and 
thus $u \simeq 2 (p_A-p_B)^2 e^{\Theta_A + \Theta_B}/(e^{\Theta_A} + e^{\Theta_B})^2$.
Using $\Theta_A = \Theta_B$, this becomes $u \simeq \frac{1}{2} (p_A-p_B)^2$ 
(also see Appendix~D in \cite{DMH11KPT}). 

\begin{remark}
Since the $p$'s are positive and distinct, for $M=2$ we have $p_A \neq p_B$ if $A \neq B$. 
For $M>2$, the latter condition holds if $p_1,\ldots,p_M$ are linearly independent over $\{ \pm 1 \}$. 
\end{remark}

\subsection{$x$-asymptotics} 
\label{subsec:x-asympt}
For fixed $t$ and suffiently large $x$ (we write $x \gg 0$), (\ref{phase_diff}) implies 
$\Theta_{(1,\ldots,1)} > \Theta_A$ 
for all $A \in \{-1,1\}^M$, $A \neq (1,\ldots,1)$. Furthermore, for $x \ll 0$ we have 
$\Theta_{(-1,\ldots,-1)} > \Theta_A$ for all $A \in \{-1,1\}^M$, 
$A \neq (-1,\ldots,-1)$. 
Hence $\cU_{(1,\ldots,1)}$ and $\cU_{(-1,\ldots,-1)}$ are dominating 
phase regions for $x \gg 0$, respectively $x \ll 0$.

\subsection{Triple phase coincidences} 
At a triple phase coincidence three different phases satisfy $\Theta_A = \Theta_B = \Theta_C$. 
If 
\bez
   p_{ABC} := p_A (q_B-q_C) + p_B (q_C-q_A) + p_C (q_A-q_B) \neq 0 \, ,  
\eez
then a corresponding event occurs at the time
\bez
    t_{ABC} := - p_{ABC}^{-1} \Big( p_A (c_B-c_C) + p_B (c_C-c_A) + p_C (c_A-c_B) \Big) \; .
\eez
Its $x$-coordinate is given by $x_{ABC} := x_{AB}(t_{ABC})$. We find that
\bez
   x_{AC}(t) - x_{AB}(t) 
 = - \frac{p_{ABC}}{(p_B-p_A)(p_C-p_A)} \, (t - t_{ABC}) \, ,
\eez
and 
\bez
   \Theta_A - \Theta_C = (p_A-p_C)(x-x_{AB}(t)) 
 - \frac{p_{ABC}}{p_B-p_A} \, (t - t_{ABC}) \, ,
\eez
so that
\be
    \Theta_A - \Theta_C = \frac{p_{ABC}}{p_A-p_B} \, (t - t_{ABC})  \qquad \mbox{on the line} 
    \quad x=x_{AB}(t) \; .     \label{phase_diff_triple}
\ee
As a consequence, the half-line
\bez
     \{ x =x_{AB}(t) \, | \, t \gtrless t_{ABC} \} \quad
     \mbox{is non-visible if} \quad \frac{p_{ABC}}{p_A-p_B}  \lessgtr 0 \; .
\eez
At a triple phase event, $\tau \simeq e^{\Theta_A} + e^{\Theta_B}+ e^{\Theta_C}$, hence 
$u \simeq \frac{4}{9} ( p_A^2 + p_B^2 + p_C^2 - p_A p_B - p_A p_C - p_B p_C)$.

\subsection{Appearances of single solitons and $t$-asymptotics} 
For $A = (\alpha_1, \ldots, \alpha_M)$, let 
$ A_{(k)} := (\alpha_1, \ldots, \alpha_{k-1}, -\alpha_k, \alpha_{k+1}, \ldots, \alpha_M)$.
On $\cU_{AA_{(k)}}$ we obtain the single soliton expression $\tau \simeq e_{-k} + e_k$, 
up to a factor that drops out in the expression for $u$. 
According to (\ref{boundary_line}), the boundary line between the two phases $\Theta_A$ and 
$\Theta_{A_{(k)}}$ is given by 
\bez
    x = x_{A A_{(k)}} 
      = - \frac{1}{2 \alpha_k p_k} (\delta_A - \delta_{A_{(k)}}) - p_k^2 \, t 
        - \sum_{j=3}^N p_k^{2j-2} \, t^{(j)}    \; .
\eez
Hence, a soliton (i.e., a visible part of such a line), moves from right to left along the $x$-axis.  
Its ``height'' is $u \simeq 2 p_k^2$. Furthermore,
\be
    x_{A A_{(k)}} - x_{B B_{(k)}}
  = - \frac{1}{2 \alpha_k p_k} (\delta_A - \delta_{A_{(k)}})
    + \frac{1}{2 \beta_k p_k} (\delta_B - \delta_{B_{(k)}})  \label{kth_soliton_line}
\ee
is constant. For $B \neq A, A_{(k)}$, the lines given by 
$x=x_{A A_{(k)}}$ and $x=x_{B B_{(k)}}$ are thus parallel. There 
are $2^{M-1}$ such lines (for fixed $k$), since this is the number of different pairs
$A, A_{(k)}$. It is natural to interprete the visible segments as appearances of the $k$th soliton. 

\begin{SCfigure}[1.8][hbtp] 
\hspace{1.cm}
\includegraphics[scale=.30]{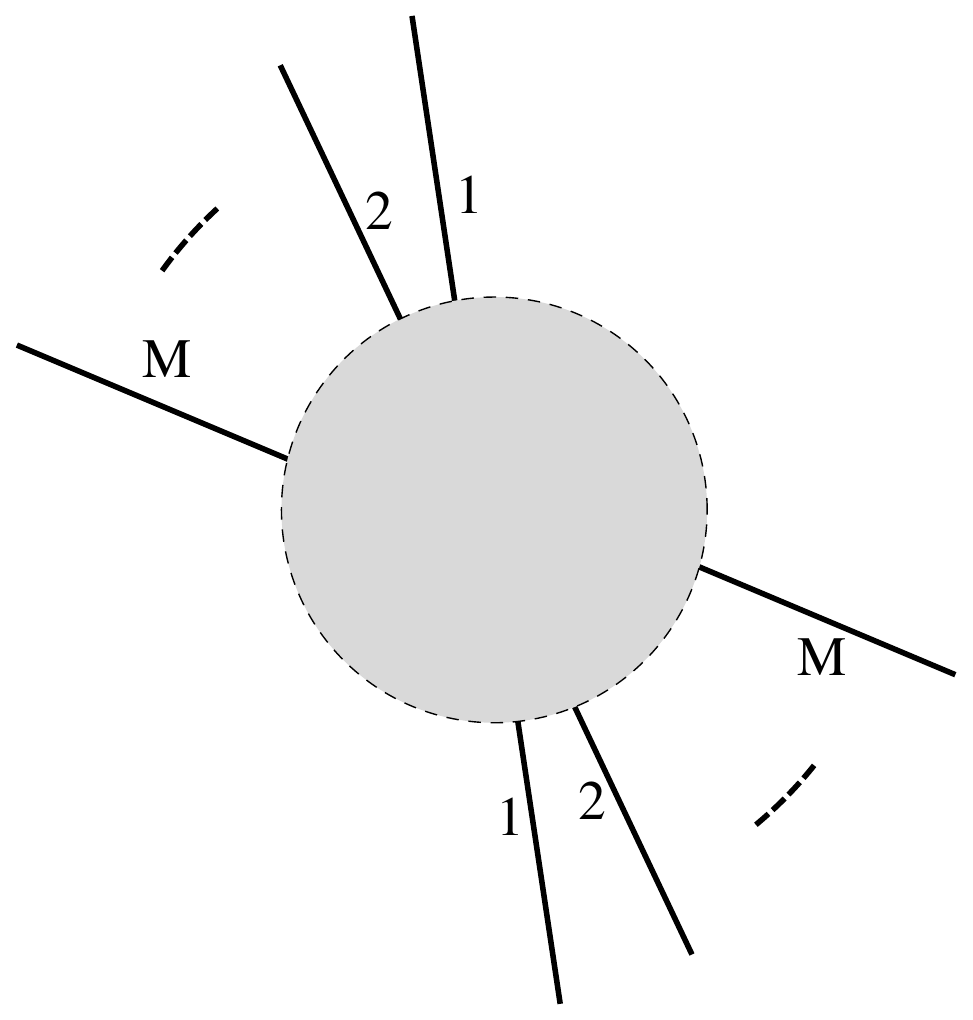} 
\caption{Asymptotic structure of an $M$-soliton KdV solution in the $xt$-plane. 
Time flows from bottom to top. A number $k$ refers to the $k$th soliton. \label{fig:KdV_asympt}  }
\end{SCfigure}

Now we show that the $k$th soliton is visible for large $|t|$.  
We consider (\ref{phase_diff_triple}) on the line $x=x_{A A_{(k)}}(t)$, with
\bez
   \frac{p_{AA_{(k)}C}}{p_A-p_{A_{(k)}}} &=& p_k^2 \, (p_C - p_A) - (q_C -q_A)
            = \sum_{j \neq k} (\gamma_j - \alpha_j) \, p_j \, (p_k^2 - p_j^2) \, , \\
            &=& \sum_{j < k} (\gamma_j - \alpha_j) \, p_j \, (p_k^2 - p_j^2) 
                + \sum_{j > k} (\alpha_j - \gamma_j) \, p_j \, (p_j^2 - p_k^2) \; .
\eez
This is different from zero if $C = (\gamma_1,\ldots,\gamma_M)$ is different from $A$ and $A_{(k)}$. 
(\ref{phase_diff_triple}) implies 
\bez
    A=(1,\ldots,1,\alpha_k,-1,\ldots,-1) \quad 
      &\Longrightarrow& \quad p_{AA_{(k)}C}/(p_A-p_{A_{(k)}}) < 0 \quad \forall C \in \{-1,1\}^M \\
      &\Longrightarrow& \quad \{ x=x_{AA_{(k)}}(t) \, | \, t \ll 0 \} \quad \mbox{is visible} \\
    A=(-1,\ldots,-1,\alpha_k,1,\ldots,1) \quad 
      &\Longrightarrow& \quad p_{AA_{(k)}C}/(p_A-p_{A_{(k)}}) > 0 \quad \forall C \in \{-1,1\}^M \\
      &\Longrightarrow& \quad \{ x=x_{AA_{(k)}}(t) \, | \, t \gg 0 \} \quad \mbox{is visible}
\eez
(using $0 < p_1 < p_2 < \cdots < p_M$). 
Moreover, for any other $A \in \{-1,1\}^M$ there is a $C \in \{-1,1\}^M$ such 
that, according to (\ref{phase_diff_triple}), $\Theta_C > \Theta_A$ for $t \ll 0$, respectively $t \gg 0$, 
so that no further boundary lines are visible for $|t| \gg 0$. 
(\ref{kth_soliton_line}) implies that
\bez
    x_{(-1,\ldots,-1) (1,-1,\ldots,-1)}
    < x_{(1,-1,\ldots,-1) (1,1,-1,\ldots,-1)} 
    < \cdots < x_{(1,\ldots,1,-1) (1,\ldots,1)}  \qquad  \mbox{for } t \ll 0 \, ,
\eez
while we have
\bez
    x_{(-1,\ldots,-1) (-1,\ldots,-1,1)}
    < x_{(-1,\ldots,-1,1) (-1,\ldots,-1,1,1)} 
    < \cdots < x_{(-1,1,\ldots,1) (1,\ldots,1)}  \qquad  \mbox{for } t \gg 0 \; .
\eez
We conclude that, as time proceeds from $-\infty$ to $+\infty$, the solitons reappear in reversed order. 
Recalling the results in Section~\ref{subsec:x-asympt}, we thus proved the asymptotic structure displayed 
in Fig.~\ref{fig:KdV_asympt}. 
In particular, this implies that a phase $\Theta_A$, where $A$ is \emph{not} of the form 
$(-1,\ldots,-1,)$, $(1,\ldots,1)$, $(1,1\ldots,1,-1,-1,\dots,-1)$ or $(-1,\ldots,-1,1,\ldots,1)$,  
can only be visible in a \emph{bounded} region of the $xt$-plane. 
Moreover, passing in the asymptotic region $|t| \gg 0$ from left to right along the $x$-axis, 
from one phase $\Theta_A$ to the next, say $\Theta_{A_{(k)}}$, requires $\alpha_k=-1$. Other 
transitions can therefore only occur in case of virtual solitons (which are bounded in space-time).

\subsection{Two solitons}
\label{sec:2solitons}
\setcounter{equation}{0}
We achieve some simplification of expressions in the following by using the notation 
$\bar{1} := -1$, and writing, e.g., $\bar{1} \ldots \bar{1} 1 \ldots 1$ instead of 
$(-1,\ldots,-1,1,\ldots,1)$. 
The tropical approximation of the $2$-soliton solution is given by
$\log \tau \simeq \max\{ \Theta_{\bar{1}\bar{1}}, \Theta_{1\bar{1}}, \Theta_{\bar{1}1}, \Theta_{11} \}$ 
with
\bez
  && \Theta_{\bar{1}\bar{1}} = - \theta_1 - \theta_2 + \log(p_2-p_1) \, , \quad
   \Theta_{1\bar{1}} = \theta_1 - \theta_2 + \log(p_1+p_2) \, , \quad  \\
  && \Theta_{\bar{1}1} = -\theta_1 + \theta_2 + \log(p_1+p_2) \, , \quad
   \Theta_{11} = \theta_1 + \theta_2 + \log(p_2-p_1) \, ,
\eez
and $\theta_j = p_j \,(x+p_j^2 t)$, $\, 0<p_1<p_2$. Here we set $N=2$. 
We recall that $\Theta_{11}$ dominates 
(all other phases) for $x \gg 0$, whereas $\Theta_{\bar{1}\bar{1}}$ dominates for $x \ll 0$. 
There are six boundary lines. Those corresponding to asymptotic solitons are
\bez
 \mbox{1st soliton:} 
    &&  x=x_{\bar{1}\bar{1},1\bar{1}}(t) = -p_1^2 t - \frac{\ell}{2 p_1}  
                    \qquad  t \ll 0 \mbox{ branch} \, , \\
    &&  x=x_{\bar{1}1,11}(t) = -p_1^2 t + \frac{\ell}{2 p_1}  
                    \qquad  t \gg 0 \mbox{ branch}           \, , \\
    &&  \mbox{shift:} \quad 
       x_{\bar{1}1,11}(t)-x_{\bar{1}\bar{1},1\bar{1}}(t) = \frac{\ell}{2p_1} 
       \qquad \mbox{where} \quad \ell := \log \frac{p_2+p_1}{p_2-p_1} > 0 \\
  \mbox{2nd soliton:}
    &&  x=x_{1\bar{1},11}(t) = -p_2^2 t + \frac{\ell}{2 p_2}  
                    \qquad  t \ll 0 \mbox{ branch} \, , \\
    &&  x=x_{\bar{1}\bar{1},\bar{1}1}(t) = -p_2^2 t - \frac{\ell}{2 p_2}  
                    \qquad  t \gg 0 \mbox{ branch} \, , \\
    &&  \mbox{shift:} \quad 
       x_{\bar{1}\bar{1},\bar{1}1}(t)-x_{1\bar{1},11}(t) = -\frac{\ell}{2p_1}
       \; .
\eez
The constant shifts exactly correspond to the well-known (asymptotically determined) \emph{phase shifts},  
which are the only witnesses of an interaction of KdV solitons. 
Further boundary lines:  
\bez
   && x = x_{\bar{1}\bar{1},11}(t) = - \frac{1}{p_1+p_2} (p_1^3 + p_2^3) \, t  
        = - (p_1^2 - p_1p_2 + p_2^2) \, t \, , \\
   && x = x_{1\bar{1},\bar{1}1}(t) = - (p_1^2 + p_1p_2 + p_2^2) \, t \; .
\eez
Triple phase coincidences occur at the times
\bez
  &&  t_{\bar{1}\bar{1},1\bar{1},\bar{1}1}
    = \frac{\ell}{2 p_1p_2 (p_2+p_1)} \, , \qquad \quad
      t_{1\bar{1},\bar{1}1,11}
    =  - \frac{\ell}{2 p_1p_2 (p_2+p_1)}  \, , \\
  &&  t_{\bar{1}\bar{1},\bar{1}1,11}
    =  - \frac{\ell}{2 p_1p_2 (p_2-p_1)} \, , \qquad \,
      t_{\bar{1}\bar{1},1\bar{1},11}
    = \frac{\ell}{2 p_1p_2 (p_2-p_1)} \; .
\eez
These times are ordered as follows: $t_{\bar{1}\bar{1},\bar{1}1,11} < t_{1\bar{1},\bar{1}1,11} < 0
  < t_{\bar{1}\bar{1},1\bar{1},\bar{1}1} < t_{\bar{1}\bar{1},1\bar{1},11}$.
Since
\bez
    \Theta_{11} - \Theta_{1\bar{1}} &=& 2 \theta_2 - \ell \\
   &=& - 2 \, \ell <0 \qquad \mbox{on} \quad 
        x=x_{\bar{1}\bar{1},\bar{1}1}(t) \, , \\
   \Theta_{11} - \Theta_{\bar{1}1} &=& 2 \theta_1 - \ell \\
   &=& - 2 \, \ell <0 \qquad \mbox{on} \quad 
        x=x_{\bar{1}\bar{1},1\bar{1}}(t) \, ,
\eez
the triple phase events at $t_{\bar{1}\bar{1},\bar{1}1,11}$
and $t_{\bar{1}\bar{1},1\bar{1},11}$ are non-visible. 
At the remaining two triple phase events 
$(x_{1\bar{1},\bar{1}1,11},t_{1\bar{1},\bar{1}1,11})$ and 
$(x_{\bar{1}\bar{1},1\bar{1},\bar{1}1},t_{\bar{1}\bar{1},1\bar{1},\bar{1}1})$, we have 
$\Theta_{11}-\Theta_{\bar{1}\bar{1}} = 2 \ell >0$, 
respectively $\Theta_{\bar{1}\bar{1}}-\Theta_{11} = 2 \ell >0$, 
so that these events are visible. Moreover, we have
\bez
     x_{1\bar{1},\bar{1}1,11} - x_{\bar{1}\bar{1},1\bar{1},\bar{1}1} 
   = \frac{p_1^2 + p_1 p_2 + p_2^2}{p_1p_2 (p_1+p_2)} \, \ell > 0 \; .
\eez

\begin{figure}[hbtp] 
\begin{center}
\includegraphics[scale=.4]{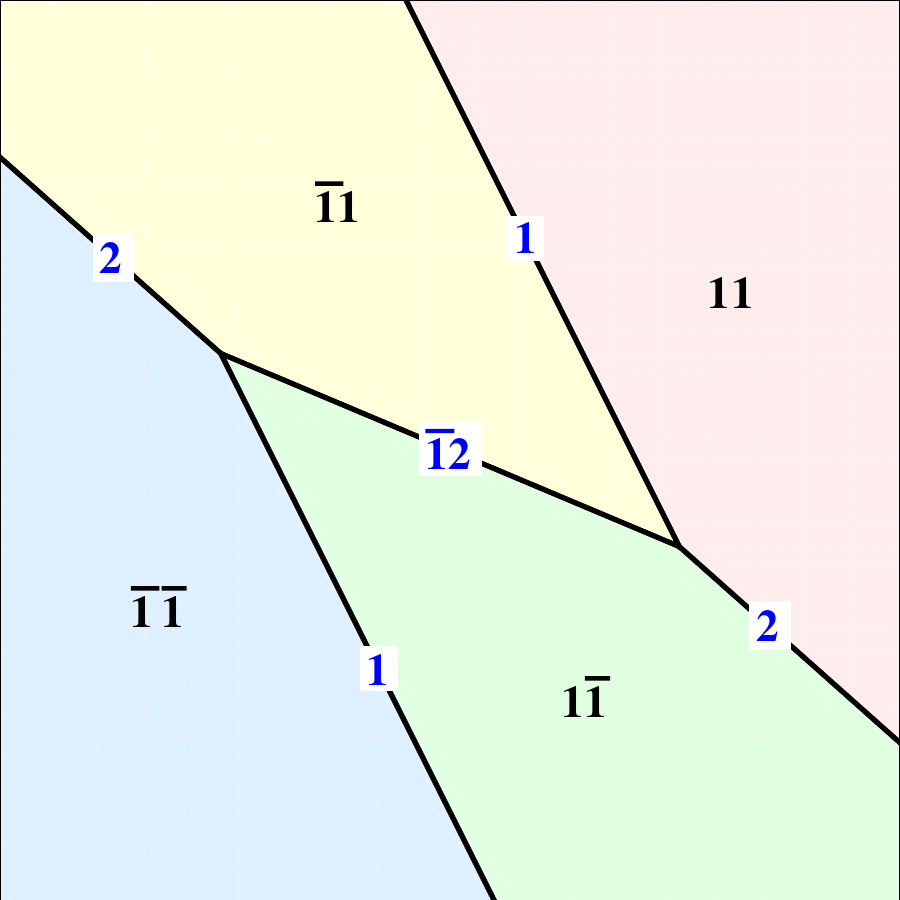}
\includegraphics[scale=.4]{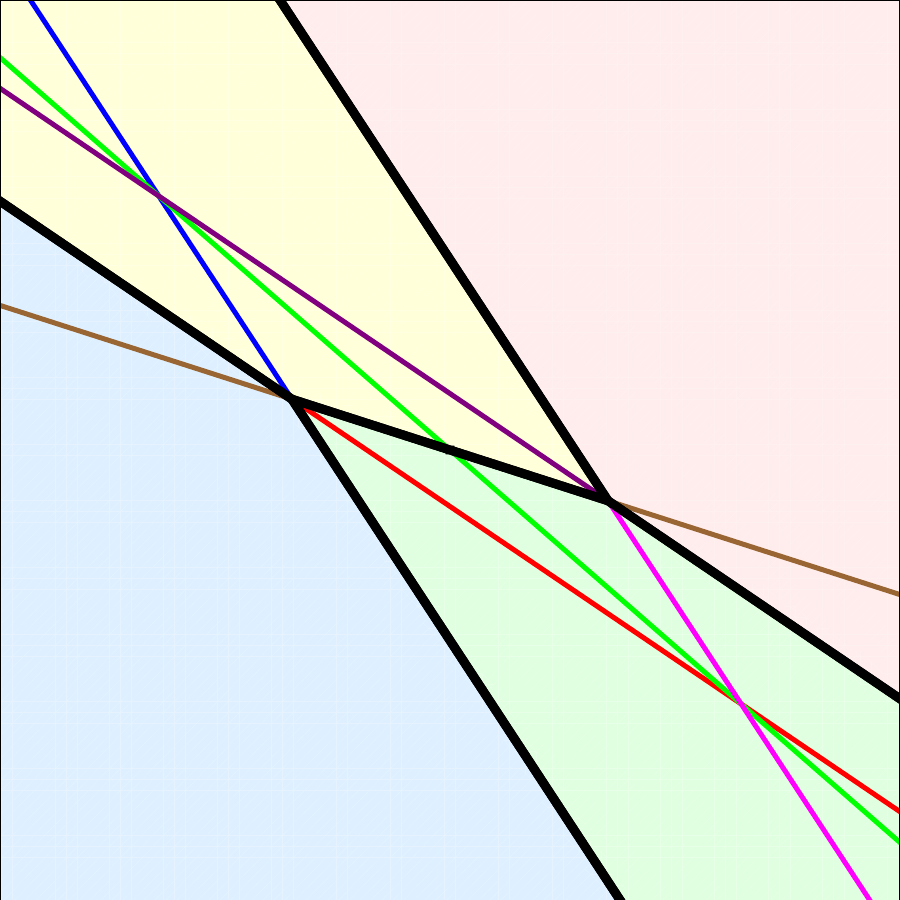}
\includegraphics[scale=.4]{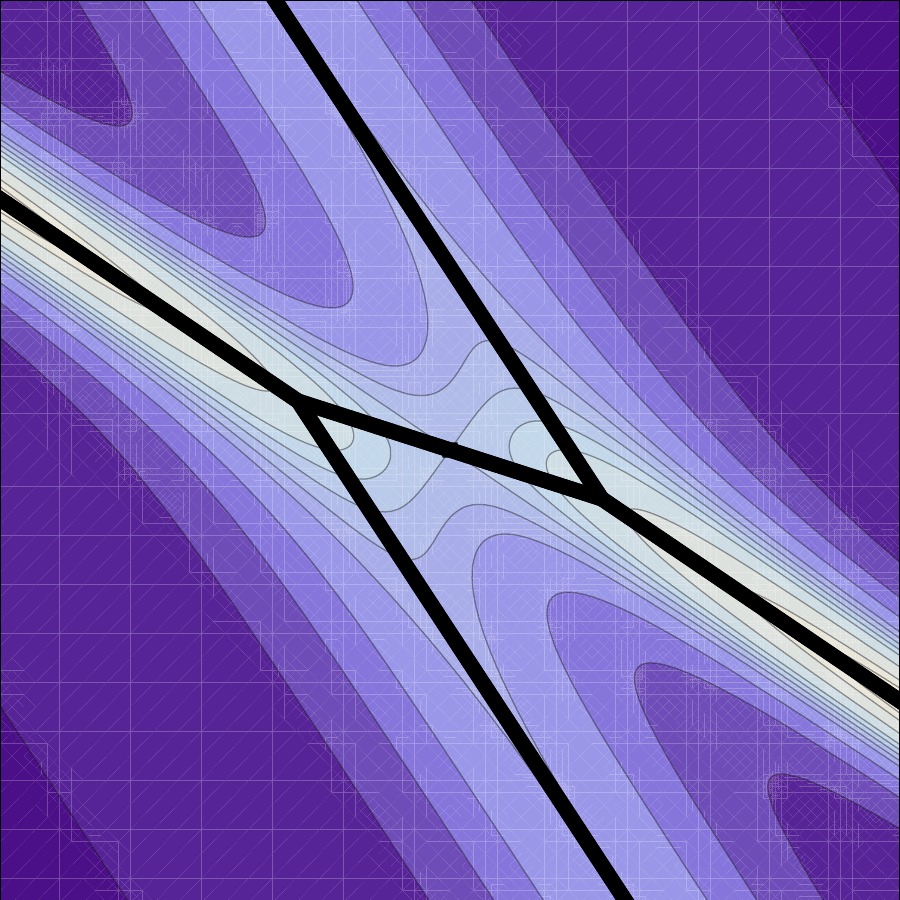}
\includegraphics[scale=.28]{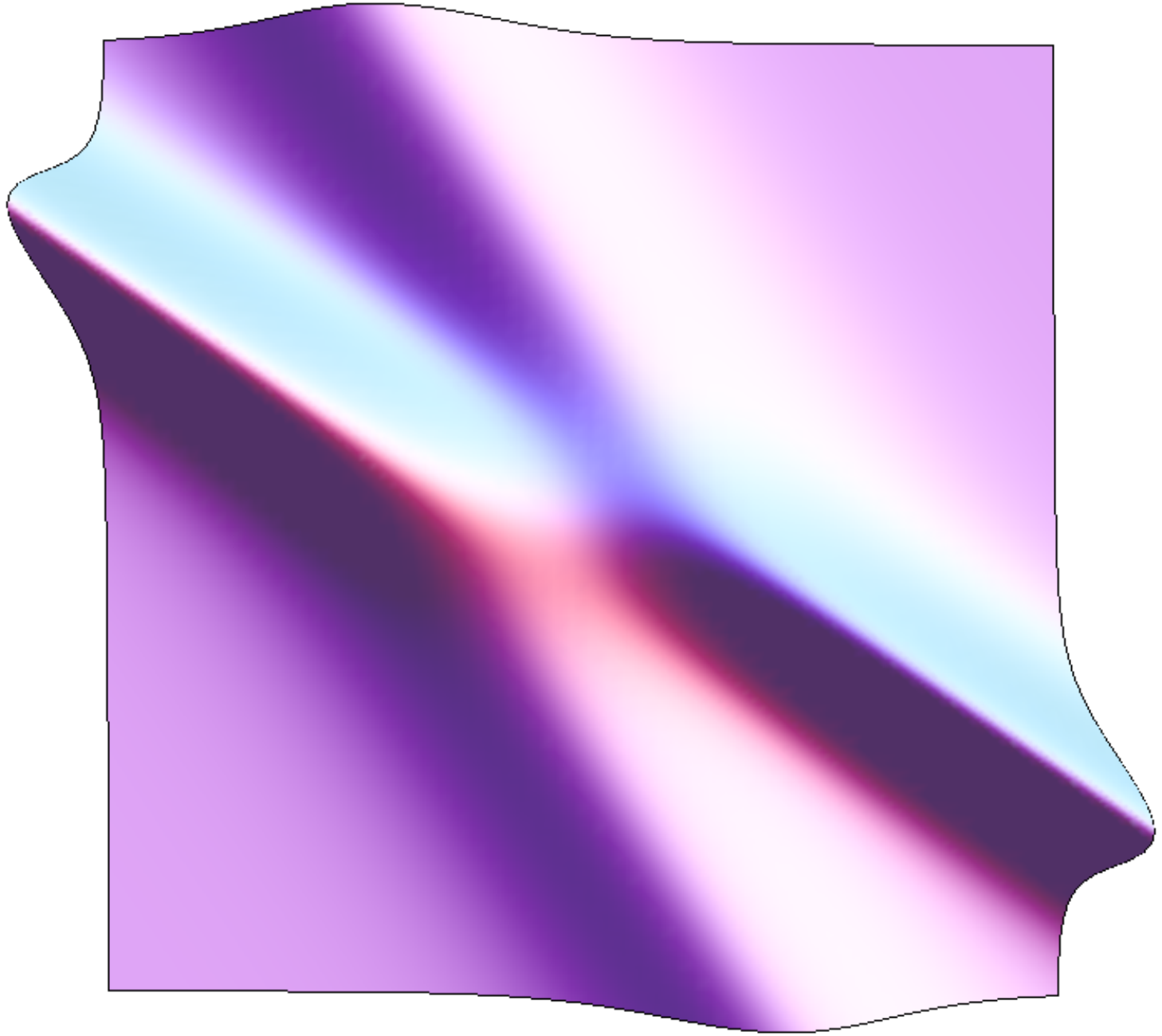}
\end{center}
\caption{To the left is a space-time plot of the tropical limit graph of a $2$-soliton KdV solution
($p_1=1, p_2=1.5$). The $x$-axis is horizontal, time proceeds from bottom to top. 
The colored regions are dominating phase regions. 
At $t= t_{1\bar{1},\bar{1}1,11}$ soliton $\boldsymbol{2}$ splits into an instance of 
soliton $\boldsymbol{1}$ and 
a ``virtual soliton'' $\boldsymbol{\bar{1}2}$, formally assigning to it the interpretation of a bound state 
composed of an antisoliton $\boldsymbol{\bar{1}}$ and a soliton $\boldsymbol{2}$. 
At $t= t_{\bar{1}\bar{1},1\bar{1},\bar{1}1}$ it merges with soliton $\boldsymbol{1}$ 
to create a new instance of soliton $\boldsymbol{2}$. 
The second plot shows all boundary lines between pairs of phases. The whole green line, 
passing through the middle point, is non-visible. 
The third plot displays the tropical limit superimposed on a contour plot of the KdV solution. 
The last is a plot of the KdV solution $u(x,t)$ over the $xt$-plane. \label{fig:KdV_2solitons_trop}  
 }
\end{figure}

Furthermore, 
\renewcommand{\arraystretch}{1.2}
\bez
    \begin{array}{l}
    \Theta_{11} - \Theta_{1\bar{1}} = \;\; 2 p_1 p_2 (p_2-p_1) \, t - \ell 
         \\
    \Theta_{11} - \Theta_{\bar{1}1} = -2 p_1 p_2 (p_2-p_1) \, t - \ell 
    \end{array}
         \qquad \mbox{on} \qquad x=x_{\bar{1}\bar{1},11}(t) \, ,
\eez
implies $\Theta_{11} < \Theta_{1\bar{1}}$ for $t<0$ and 
$\Theta_{11} < \Theta_{\bar{1}1}$ for $t>0$ on the boundary line $x=x_{\bar{1}\bar{1},11}(t)$, 
so that the whole line is non-visible. 
We thus arrive at the situation described in Fig.~\ref{fig:KdV_2solitons_trop}.
Two solitons interact by exchanging a ``virtual soliton''. 

\subsection{Three solitons}
\begin{SCfigure}[2.8][hbtp]
\hspace{.6cm}
\caption{
According to Remark~\ref{rem:reflection}, $u(x,t,s)$ has the 
property $u(x,0,0)=u(-x,0,0)$. Hence $u(x,0,0)$ has an extremum at $x=0$, 
determined by the sign of 
$\frac{1}{4} u_{xx}(0,0,0) = - (p_1^4 + 3 p_2^4 + p_3^4) + 4 (p_1^2 p_2^2 - p_1^2 p_3^2 + p_2^2 p_3^2)$. 
The plot displays for $p_3/p_2$ (vertical axis) versus $p_2/p_1$ (horizontal axis) the boundary 
between the regions where $u(x,0,0)$ has a minimum (light region), respectively 
a maximum (dark region).   \label{fig:3solitons_min-max}
}
\includegraphics[scale=.36]{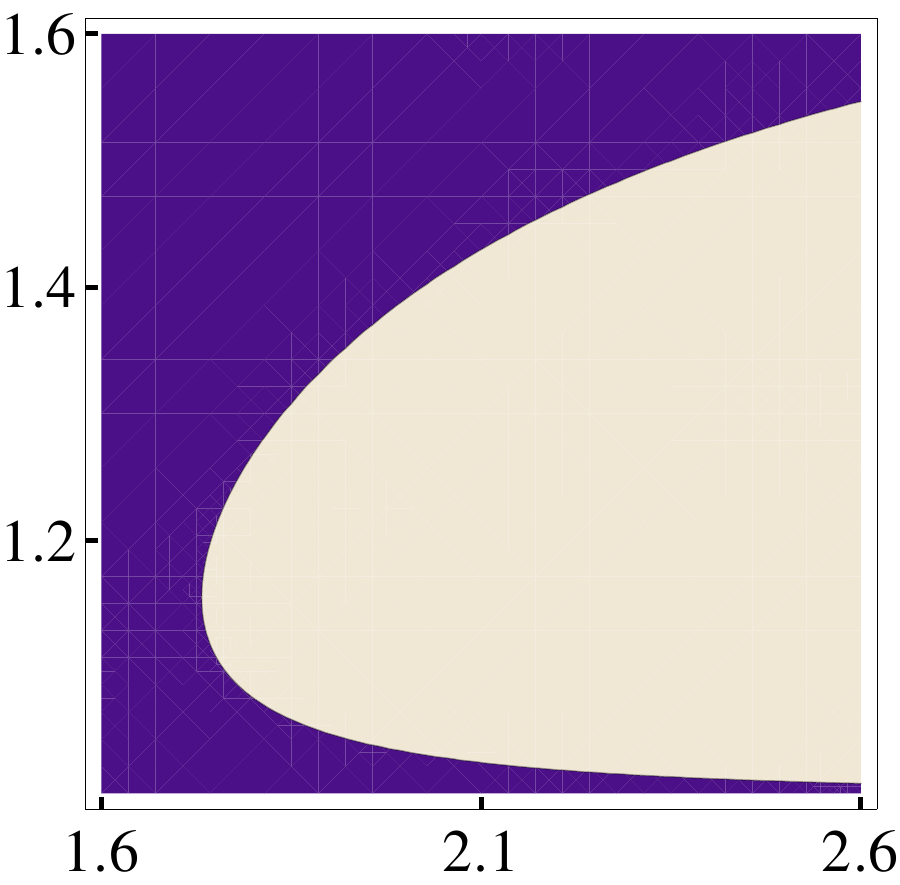} 
\end{SCfigure}

\begin{figure}[hbtp]
\hspace{.1cm}
\begin{center}
\includegraphics[scale=.4]{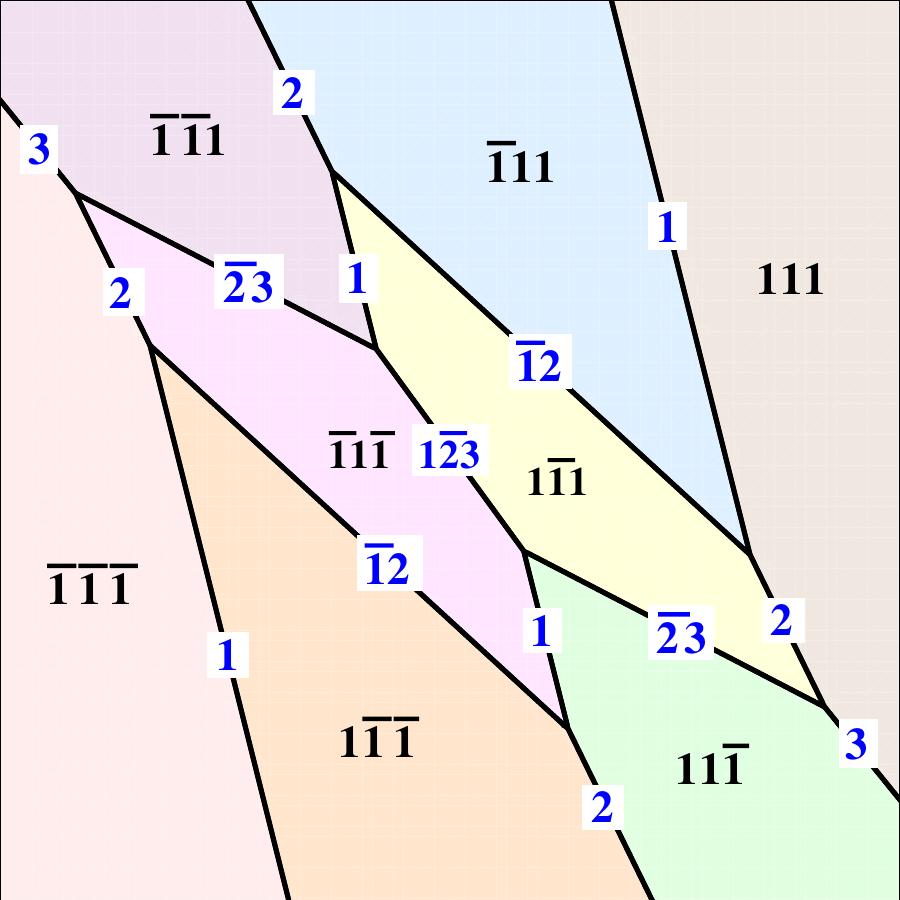} 
\hspace{.3cm}
\includegraphics[scale=.244]{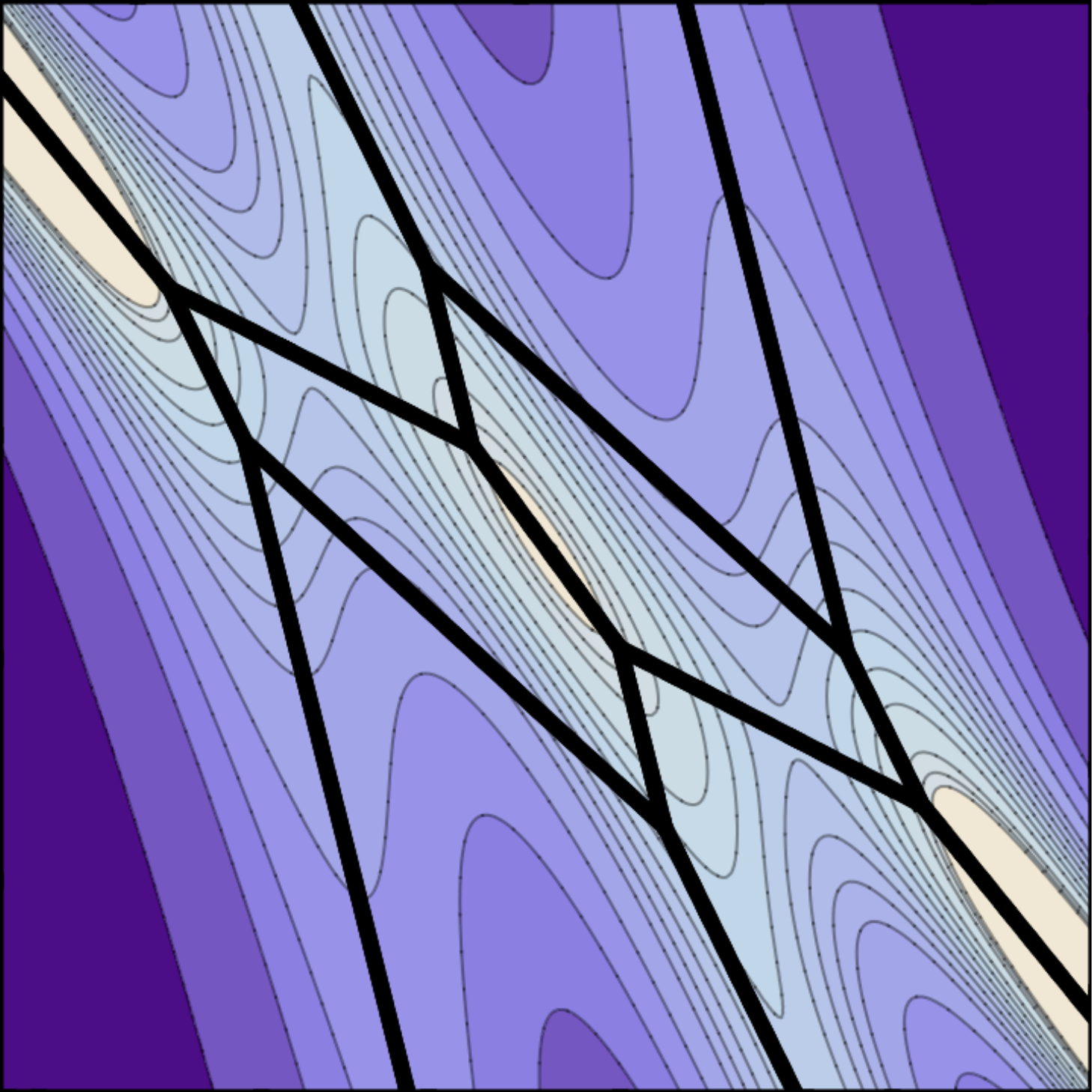} 
\hspace{.3cm}
\includegraphics[scale=.252]{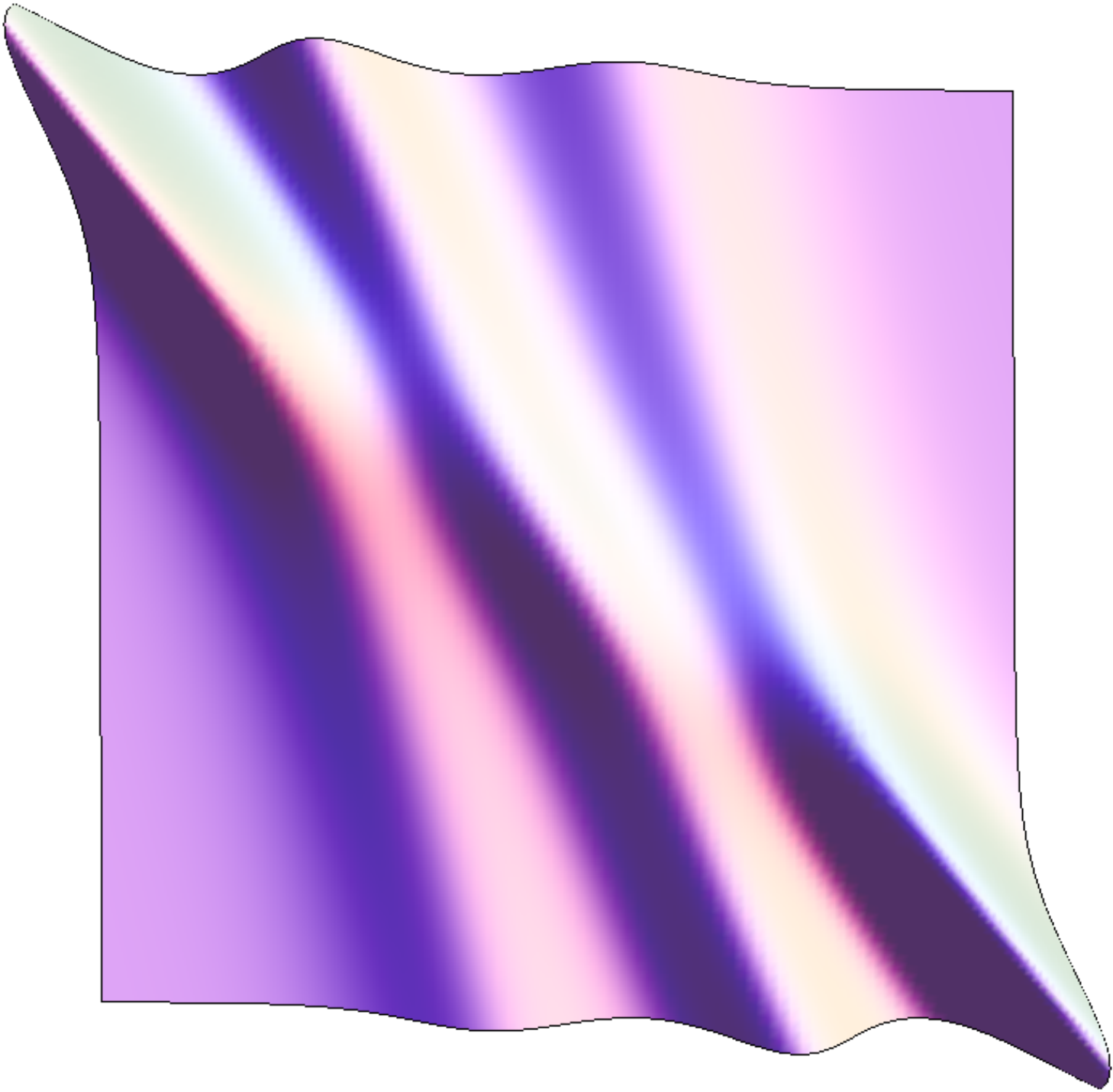} 
\end{center}
\caption{The first plot displays the tropical limit of a $3$-soliton KdV solution for $s=0$
(and $p_1=0.5$, $p_2=0.7$, $p_3=0.9$). 
Here all $2^3$ phases are visible. The three phase regions extending to the 
bottom ($t \ll 0$) are given, from left to right, by  
$\Theta_{\bar{1}\bar{1}\bar{1}}, \Theta_{1\bar{1}\bar{1}}, \Theta_{11\bar{1}}$, 
those extending to the top ($t \gg 0$) by 
$\Theta_{\bar{1}\bar{1}1}, \Theta_{\bar{1}11}, \Theta_{111}$. 
In the middle we have two bounded phase regions where $\Theta_{\bar{1}1\bar{1}}$, respectively
$\Theta_{1\bar{1}1}$, dominate.
The second picture shows the superimposition of the tropical limit on 
a contour plot of the KdV solution. The third is a plot of the KdV solution $u(x,t,s)$ 
at $s=0$.
\label{fig:tropicalKdV_3solitons} }
\end{figure}

\setcounter{equation}{0}
Setting $N=3$, we now have $\theta_j = p_j \, ( x + p_j^2 \, t + p_j^4 \, s)$,
$j=1,2,3$, with $s := t^{(3)}$. Recall that $0 < p_1 < p_2 < p_3$. 
The tropical approximation of the $3$-soliton solution is given by
$ \log \tau \simeq \max\{ \Theta_0, \Theta_1, \ldots, \Theta_7 \} $,
where $\Theta_0 := \Theta_{\bar{1}\bar{1}\bar{1}}$, $\Theta_1 := \Theta_{1\bar{1}\bar{1}}$, 
$\Theta_2 := \Theta_{\bar{1}1\bar{1}}$, $\Theta_3 := \Theta_{11\bar{1}}$, 
$\Theta_4 := \Theta_{\bar{1}\bar{1}1}$, $\Theta_5 := \Theta_{1\bar{1}1}$, 
$\Theta_6 := \Theta_{\bar{1}11}$, $\Theta_7 := \Theta_{111}$.
Only the $\Theta_2$- and $\Theta_5$-regions are \emph{bounded} in the $xt$-plane. 
In the following, we present results derived with the help of computer algebra. 
It turns out that there are no visible coincidences of more than four of these 
phases. 
Among the ${8 \choose 4} = 70$ a priori possible coincidences of four phases, 
the following can be ruled out: $\{0,1,2,3\}$, $\{0,1,4,5\}$, $\{0,1,6,7\}$, $\{0,2,4,6\}$, 
$\{0,2,5,7\}$, $\{0,3,4,7\}$, $\{1,2,5,6\}$, $\{1,3,4,6\}$, $\{1,3,5,7\}$, $\{2,3,4,5\}$, 
$\{2,3,6,7\}$, $\{4,5,6,7\}$, where a number $i$ represents the phase $\Theta_i$. 
Of the remaining 58 4-phase coincidences, most are non-visible (for all values of the parameters). 
They can only occur if $s=s_{ijkl}$, where $s_{ijkl}$ is completely determined in 
terms of the parameters $p_m$:
\bez
    s_{ijkl} &=& \frac{a_{ijkl}}{2 p_2 (p_2^2-p_1^2)(p_3^2-p_2^2)}
                 + \frac{b_{ijkl}}{2 p_1 (p_2^2-p_1^2)(p_3^2-p_1^2)} 
                 + \frac{c_{ijkl}}{2\, p_3 (p_3^2-p_2^2)(p_3^2-p_1^2)} \, ,
\eez
where
\bez
\begin{array}{l@{\qquad}l@{\qquad}l}
 a_{0124} = \ell_{12} + \ell_{23} &  b_{0124} = -\ell_{12} - \ell_{13} & c_{0124} = - \ell_{13} - \ell_{23} \\
 a_{1235} = \ell_{23} - \ell_{12} &  b_{1235} = \ell_{12} - \ell_{13} & c_{1235} =  \ell_{13} - \ell_{23} \\
 a_{1245} = -2 \ell_{13} + \ell_{12} + \ell_{23} & b_{1245} = \ell_{13} - \ell_{12} 
          & c_{1245} = \ell_{13} - \ell_{23} 
\end{array}
\eez
with $\ell_{ij} := \log[(p_j+p_i)/(p_j-p_i)]$, and we find 
\bez 
    s_{2356} = - s_{1245} \, , \qquad
    s_{2456} = - s_{1235} \, , \qquad
    s_{3567} = - s_{0124} \; .
\eez
A 4-phase coincidence described by $\{i,j,k,l\}$ can only be visible if for each subset of 
three indices the corresponding 3-phase coincidence is visible. Table~\ref{table:3phase_coinc}
lists all (under the stated conditions) visible 3-phase coincidences. 
Fig.~\ref{fig:3solitons_min-max} provides some more information about the $3$-soliton case 
and Fig.~\ref{fig:tropicalKdV_3solitons} shows an example.

\begin{table}
\caption{Results about 3-phase coincidences for $3$-soliton solutions. \label{table:3phase_coinc}}
\begin{center}
\begin{tabular}{l|cl}
 3-phase coincidence time & visibility condition &   \\ 
\hline
 $t_{012}$, $t_{024}$ & $s < s_{0124}$ &             \\
 $t_{014}$ & $s_{0124} < s$ &                            \\
 $t_{124}$ & $s_{0124} < s < s_{1245}$     &             \\
 $t_{145}$ & $s_{1245} < s$ &                            \\
 $t_{245}$ & $s_{2456} < s < s_{1245}$     &             \\
 $t_{125}$ & $s_{1235} < s < s_{1245}$     & if $\; p_3 < \sqrt{p_1^2+p_1p_2+p_2^2}$  \\
         & $s_{1245} < s < s_{1235}$     & if $\; p_3 > \sqrt{p_1^2+p_1p_2+p_2^2}$  \\
 $t_{123}$ & $s < s_{1235}$ &                            \\
 $t_{135}$ & $s_{1235} < s$ &                            \\
 $t_{235}$ & $s_{2356} < s < s_{1235}$     &             \\
 $t_{236}$ & $s < s_{2356}$ &                            \\
 $t_{256}$ & $s_{2356} < s < s_{2456}$ & if $\; p_3 < \sqrt{p_1^2+p_1p_2+p_2^2}$     \\
           & $s_{2456} < s < s_{2356}$ & if $\; p_3 > \sqrt{p_1^2+p_1p_2+p_2^2}$     \\           
 $t_{356}$ & $s_{2356} < s < s_{3567}$ &                 \\
 $t_{357}$, $t_{367}$, $t_{567}$  & $s_{3567} < s$ &  \\
 $t_{456}$ & $s_{2456} < s$ &                                 \\
 $t_{246}$ & $s < s_{2456}$ & 
\end{tabular}
\end{center}
\end{table}
Except for some relations implied by Table~\ref{table:3phase_coinc}, like 
$s_{0124} < s_{1245}$, the order of the values $s_{ijkl}$ depends in a more complicated way 
on the parameters $p_j$. In particular, we can draw the following conclusions. 
\begin{itemize}
\item Only the 4-phase coincidences corresponding 
to $\{0,1,2,4\}$, $\{1,2,3,5\}$, $\{1,2,4,5\}$, $\{2,3,5,6\}$, 
$\{2,4,5,6\}$, $\{3,5,6,7\}$ are visible, for certain values of $s$.
\item $\Theta_2$ is non-visible if either $s > s_{1245}$ and $p_3 < \sqrt{p_1^2+p_1p_2+p_2^2}$, 
or $s > s_{1235}$ and $p_3 > \sqrt{p_1^2+p_1p_2+p_2^2}$ (since then all 3-phase coincidences
with $t_{ijk}$, $2 \in \{i,j,k\}$, are non-visible). In particular, $\Theta_2$ is non-visible 
if $s \gg 0$.  
\item $\Theta_5$ is non-visible if either $s < s_{2356}$ and $p_3 < \sqrt{p_1^2+p_1p_2+p_2^2}$, 
or $s < s_{2456}$ and $p_3 > \sqrt{p_1^2+p_1p_2+p_2^2}$. In particular, $\Theta_5$ is non-visible if 
$s \ll 0$. 
\end{itemize}

Although some general results can be obtained about arbitrarily high order phase coincidences, 
there is hardly a chance to achieve a classification, for arbitrary $M$, of all possible 
evolutions of $M$-soliton KdV solutions or, equivalently, parallel $M$-soliton KP-II solutions,  
in a similar way as for the case of tree-shaped KP line soliton solutions \cite{DMH11KPT,DMH12KPBT}. 
The simple combinatorics (higher Tamari orders), underlying the latter case, has  
no counterpart in case of more general KP line soliton solutions (also see, e.g., \cite{Koda+Will11} 
for an analysis of the general KP case).

F M-H is grateful to Boris Konopelchenko for a very stimulating discussion about tropical and  
dispersionless limits.

\end{document}